\documentclass[aps,prl,twocolumn,groupedaddress,showpacs]{revtex4}
\usepackage{graphics}
\usepackage{amsmath}
\usepackage{epsfig}
\usepackage{epsf}
\newcommand {\sla}[1]{ #1 \!\!\!/}

\begin{document}
\title{The Two-Photon-Exchange and $\gamma Z$-Exchange Corrections to \\Parity-Violating Elastic Electron-Proton Scattering}

\date{\today}

\author{ Hai Qing Zhou$^{1*}$,
Chung Wen Kao$^2$, and Shin Nan Yang$^{1,3}$ \\
$^1$Department of Physics and $^3$Center for Theoretical Sciences,
National Taiwan University,\\
Taipei 10617, Taiwan,\\
$^2$Department of Physics, Chung-Yuan Christian University,\\
Chung-Li 32023, Taiwan\\
}
\begin{abstract}
Leading electroweak corrections play an important role in precision
measurements of the strange form factors. We calculate the
two-photon-exchange (TPE) and $\gamma Z$-exchange corrections to the
parity-violating asymmetry of the elastic electron-proton scattering
in a simple hadronic model including the finite size of the proton.
We find both can reach a few percent and are comparable in size with
the current experimental measurements of strange-quark effects in
the proton neutral weak current. The effect of $\gamma Z$-exchange
is in general larger than that of TPE, especially at low momentum
transfer $Q^2\le 1 GeV^2$. Their combined effects on the values of
$G^s_E+\beta G^s_M$ extracted in recent experiments can be as large
as $-40\%$ in certain kinematics.

\end{abstract}
 \maketitle

Strangeness content in the proton remains one of the most
intriguing questions in hadron structure. Early indications on the
contribution of strange quarks to the nucleon properties came from
neutrino and electron deep inelastic scatterings and pion-nucleon
sigma term, which suggested that strange quarks might give
non-negligible contributions to the spin and mass of the proton
\cite{Beck01}. Many other observables were later suggested,
including excess $\phi$ production in $p\bar p$ annihilation
\cite{Amsler98}, double polarizations in photo- and
electroproduction of $\phi$ meson \cite{Titov97}, and asymmetry in
scattering of longitudinally polarized electrons from polarized
targets, to probe the strangeness in the nucleon.

Parity-violating asymmetry
$A_{PV}=(\sigma_R-\sigma_L)/(\sigma_R+\sigma_L)$ in polarized
electron elastic scattering arises from the interference of weak
and electromagnetic amplitudes. Weak neutral current elastic
scattering is mediated by the $Z$-exchange and measures form
factors which are sensitive to a different linear combination of
the three light quark distributions.  When combined with proton
and neutron electromagnetic form factors and with the use of
charge symmetry, the strange electric and magnetic form factors,
$G^s_E$ and $G^s_M$, can then be determined \cite{Kaplan}. Since
this is a rather clean technique to access the charge and
magnetization distributions of the strange quark within nucleons,
four experimental programs SAMPLE \cite{SAMPLE}, HAPPEX
\cite{HAPPEX}, A4 \cite{A4}, and G0 \cite{G0} have been designed
to measure this important quantity, which is small and ranges from
0.1 to 100 ppm.
This calls for greater
efforts to reduce theoretical uncertainty in order to arrive at a
more reliable interpretation of experiments.

\begin{figure}[t]
\centerline{\epsfxsize 1.4 truein\epsfbox{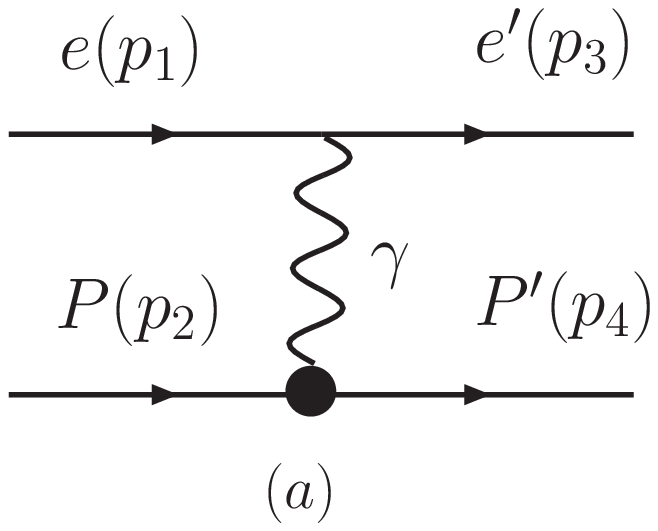}
\epsfxsize 1.4 truein\epsfbox{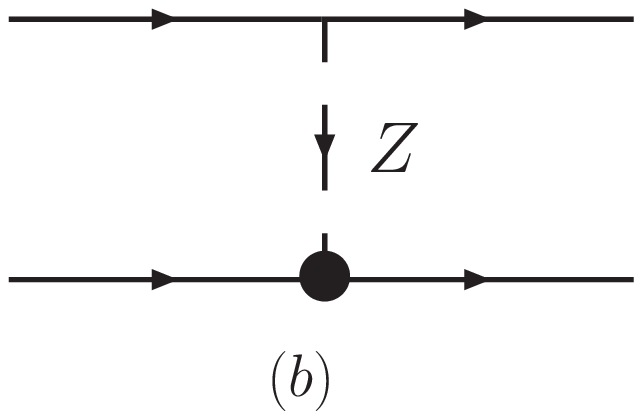}} \centerline{\epsfxsize
1.4truein\epsfbox{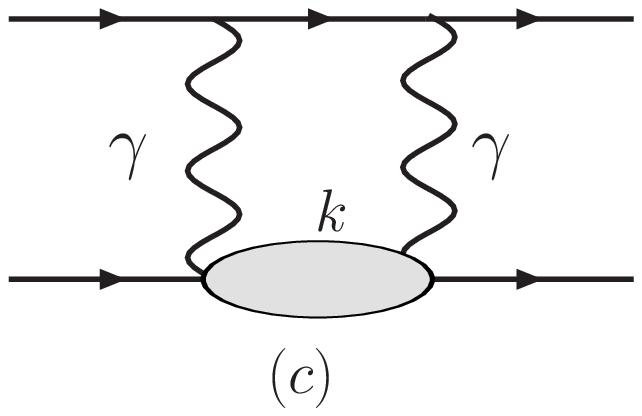} \epsfxsize 1.4
truein\epsfbox{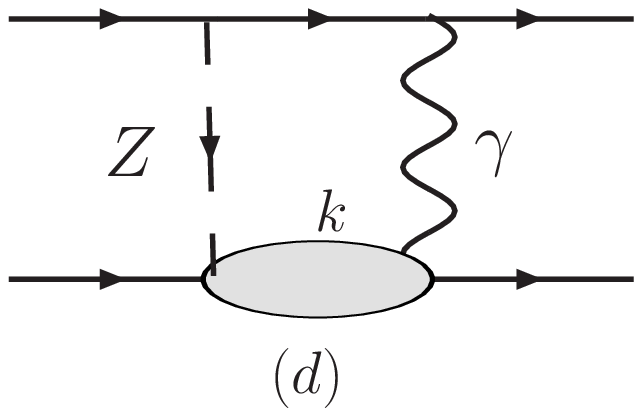}} \caption{(a) one-photon-exchange,
(b) $Z$-boson-exchange, (c) TPE, and (d) $\gamma Z$-exchange
diagrams for elastic {\it ep} scattering. Corresponding cross-box
diagrams are implied.}
\end{figure}

At tree level, parity violation in electron scattering
$e(p_1)+p(p_2)\rightarrow e(p_3)+p(p_4)$ comes from the
interference of diagrams with one-photon-exchange (OPE) and
$Z$-boson exchange shown, respectively, in Figs. 1(a) and 1(b).
Leading order radiative corrections include the box diagrams shown
in Figs. 1(c) and 1(d) and other diagrams. The radiative
corrections to $A_{PV}$ have been discussed in
\cite{Marciano83,Musolf92}. However, theoretical uncertainties
remain.

Recently, the contribution of the interference of the
two-photon-exchange (TPE) process of Fig. 1(c) with diagram of Figs.
1(a) and 1(b) to $A_{PV}$, has been evaluated in \cite{Afanasev05}
in a parton model using GPDs. It was prompted by the fact that such
a model calculation of the TPE effect was found \cite{Chen04} to be
arguably able to explain  the discrepancy between the measurement of
the proton electric to magnetic form factor ratio $R=\mu_pG_E/G_M$,
where $\mu_p=2.79$, from Rosenbluth technique and polarization
transfer technique at high momentum-transfer-squared $Q^2$
\cite{Jone00}. In \cite{Afanasev05}, it was found that indeed the
TPE correction to $A_{PV}$ can reach several percent in certain
kinematics, becoming comparable in size with existing experimental
measurements of strange-quark effects in the proton neutral weak
current. However, the partonic calculations of
\cite{Afanasev05,Chen04} are reliable only for $Q^2$ large
comparable to a typical hadronic scale, while all current
experiments \cite{SAMPLE,HAPPEX,A4,G0} have been performed at lower
$Q^2$ values. In addition, the $\gamma Z$-exchange diagram of Fig.
1(d), expected to be of the same order as the TPE correction, was
not considered in \cite{Afanasev05}.

In this paper, we report on calculations of the TPE and $\gamma
Z$-exchange corrections to $A_{PV}$, where both are treated in the
same hadronic model developed in \cite{Blunden03} to estimate the
TPE contribution to $R$. The advantage of the calculation of
\cite{Blunden03} is that it is also applicable to low $Q^2$ region
and their results for $R$ are in agreement with the partonic
calculation of \cite{Chen04}. We will follow \cite{Blunden03} and
consider only the elastic intermediate states in the blobs of
Figs. 1(c) and 1(d). We find both can reach a few percent and are
comparable in size with the current experimental constraints of
strange-quark effects in the proton neutral weak current.

At hadron level, the couplings of photon and $Z$-boson with proton
are given as
\begin{eqnarray}
\langle p'|J^Z_\mu|p\rangle
&=&\overline{u}(p')[F^{Z,p}_{1}\gamma_\mu+F^{Z,p}_{2}\frac{i\sigma_{\mu\nu}}{2M}q^\nu+G^Z_A
\gamma_\mu\gamma_5]u(p), \nonumber \\
\langle p'|J^\gamma_\mu|p\rangle
&=&\overline{u}(p')[F^{\gamma,p}_{1}\gamma_\mu+F^{\gamma,p}_{2}
\frac{i\sigma_{\mu\nu}}{2M}q^\nu]u(p),
\end{eqnarray}
where $M$ is the proton mass and $q=p'-p$. $F^{\gamma/Z,p}_{1,2}$
and $G^Z_A$ are the proton electromagnetic/neutral weak current
and axial form factors, respectively.

Choosing the Feynman gauge and neglecting the electron mass $m_e$ in
the numerators, the amplitudes of box diagrams Fig. 1(c) and Fig.
1(d) can be written as
\begin{align}
&M^{(d)}=-i\int\frac{d^4k}{(2\pi)^4}\overline{u}(p_3)(-ie\gamma^{\mu})\frac{i(\sla{p}_1+\sla{p}_2-\sla{k})}
{(p_1+p_2-k)^2-m_e^2+i\varepsilon} \notag \\
&\times(-ig\gamma^\nu)((-1+4\sin^{2}\theta_{W})+\gamma_{5})u(p_1)
\frac{-i}{(p_4-k)^2-\lambda^2+i\varepsilon}\notag \\
&\times\frac{-i}{(k-p_2)^2-M_Z^2+i\varepsilon}
\overline{u}(p_4)\Gamma^\gamma_{\mu}\frac{i(\sla{k}+M)}
{k^2-M^2+i\varepsilon}\Gamma^Z_\nu u(p_2),\notag \\
&M^{(c)}=-i\int\frac{d^4k}{(2\pi)^4}\overline{u}(p_3)(-ie\gamma^{\mu})\frac{i(\sla{p}_1+\sla{p}_2-\sla{k})}
{(p_1+p_2-k)^2-m_e^2+i\varepsilon}\notag\\
&\times(-ie\gamma^{\nu})
u(p_1)\frac{-i}{(p_4-k)^2-\lambda^2+i\varepsilon}\frac{-i}{(k-p_2)^2-\lambda^2+i\varepsilon}\notag \\
&\times\overline{u}(p_4) \Gamma^\gamma_{\mu}\frac{i(\sla{k}+M)}
{k^2-M^2+i\varepsilon}\Gamma^\gamma_{\nu}u(p_2),\label{eq:diagram}
\end{align}

\noindent where $\Gamma^{\gamma}_\mu$=$ie\langle
p'|J^\gamma_\mu|p\rangle$ and $\Gamma^Z_\mu$=$-ig\langle
p'|J^Z_\mu|p\rangle$ with $4g^2/M_{Z}^2=\sqrt{2}G_F$,  $M_Z$ the
$Z$-boson mass, and $G_F$ the Fermi constant. Amplitudes for the
cross-box diagrams can be written down similarly. The
infinitesimal photon mass $\lambda$ has been introduced in the
photon propagator to regulate the IR divergence. In the soft
photon approximation, the sums of $M^{(c)}$, $M^{(d)}$ and their
crossed diagrams can be factorized as
\begin{equation}
M^{(c)+(c')}_{soft}=\frac{1}{2}\delta_{MT}M^{(a)},\,\,\,
M^{(d)+(d')}_{soft}=\frac{1}{2}\delta_{MT}M^{(b)},
\end{equation}
where $\delta_{MT}$
denotes the correction from the box diagrams in the soft photon
approximation given by the standard treatment of Mo and Tsai
\cite{MoTsai}. The IR divergence of the interference of
$M^{(c)+(c')}_{soft}$ and $M_{soft}^{(d)+(d')}$ with Figs. 1(a)
and 1(b) are exactly canceled by corresponding terms in the
bremsstrahlung cross section involving the interference between
real photon emission from the electron and from the proton. Under
such an approximation, the box diagrams and their corresponding
bremsstrahlung cross section give no correction to $A_{PV}$ since
$\delta_{MT}$ is independent of the initial electron helicity. To
go beyond the soft photon approximation to estimate the
corrections to $A_{PV}$, we calculate the full amplitudes of
$M^{(c)}$ and $M^{(d)}$ and subtract $M^{(c)}_{soft}$ and
$M^{(d)}_{soft}$ from their respective full amplitude. The
interferences between the remaining box diagrams and the tree
diagrams are thus IR safe.

To calculate the full amplitudes of $M^{(c)}$ and $M^{(d)}$, we
need explicit forms of the form factors. For simplicity, we choose
to parameterize the Sachs form factors of the proton
$G^{\gamma/Z,p}_{E}$ and $G^{\gamma/Z,p}_{M}$, which are linear
functions of $F^{\gamma/Z,p}_{1}$ and $F^{\gamma/Z,p}_{2}$, in
monopole forms: $G^{\gamma,p}_{E} = G^{\gamma,p}_{M}/\mu_p =
G^{Z,p}_{E}/x = G^{Z,p}_{M}/y =   \Lambda^2_1/(Q^2+\Lambda^2_1)$,
$G^Z_A/z =
 \Lambda^2_2/(Q^2+\Lambda^2_2)$.  Here we assume that
$\mu_pG^{\gamma,p}_{E}/G^{\gamma,p}_{M} =1.$ Actually we find that
if one assumes $\mu_pG^{\gamma,p}_{E}/G^{\gamma,p}_{M} =
\Lambda_3^2/(Q^2+\Lambda_3^2)$, the results are insensitive to the
value of $\Lambda_3$ when $\Lambda_3\ge 2\, GeV$. We take $\Lambda_1
= 0.56\, GeV, \Lambda_2 = 0.7 \,GeV$ by fitting to the usual dipole
forms \cite{HAPPEX,Beise} $G^{\gamma,p}_E = 1/(1+Q^2/0.71)^2$,
$G^Z_A = G^Z_A(0)/(1+Q^2)^2$, with $Q$ given in unit of $GeV$, i.e.,
$c=1$, a convention to be used hereafter. $x, y, z$ are determined
from relations \cite{Beise}, $G^{Z,p}_{E,M} = \rho(1-4\kappa
\sin^2\theta_W)G^{\gamma,p}_{E,M}-\rho G^s_{E,M} -\rho
G^{\gamma,n}_{E,M}$ and $G^Z_A = -(1+R_A^{T=1})G_A+\sqrt{3}R_A^{T=0}
G_A^8+\Delta s$ at $Q^2=0$ point. The quantities $G_A^8$ and $\Delta
s$ refer to the $SU(3)$ isoscalar octet form factor and the strange
quark contrbution to the nucleon spin, respectively. The $\rho,
\kappa$ and $R_{A}^{T=1}$ and $R_{A}^{T=0}$ are due to radiative
corrections. This results in $x=0.076\pm0.00264$, $y=2.08 \pm
0.00813-G_M^s(0), z = -0.95^{+0.37}_{-0.36}+\Delta s(0)$. We fix $x
= 0.076$ and vary the values of $y, z, \Lambda_1,$ and $\Lambda_2$
to check the sensitivity of the results on the parameters and get
almost the same results.  As in \cite{Blunden03}, we use package
FeynCalc~\cite{Feyncalc} and  LoopTools~\cite{Looptools} to do the
analytical and numerical calculations, respectively. The IR
divergence has been checked in our calculation.

\begin{figure}[h,b,t]
\centerline{\epsfxsize 4.0 truein\epsfbox{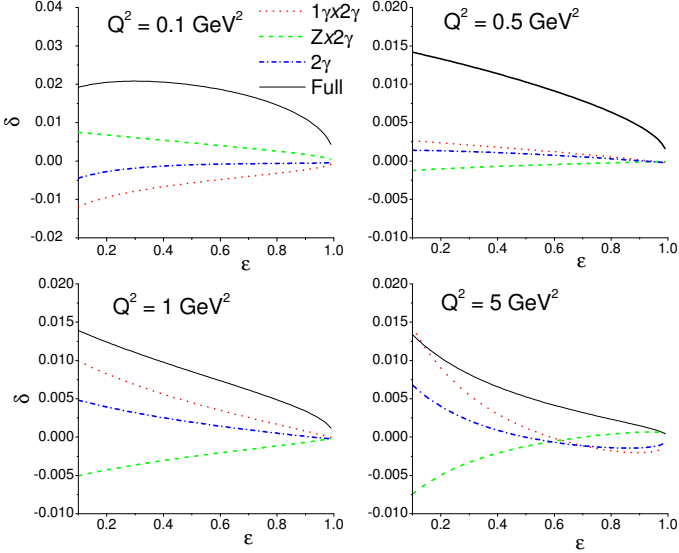}}
\caption{TPE and $\gamma$Z-exchange corrections to parity-violating
asymmetry as functions of $\epsilon$ from 0.1 to 0.99 at $Q^2=0.1,\,
0.5,\, 1.0,$ and $5.0\, GeV^2$. Dotted and dashed lines denote
correction coming only from the interference between Fig. 1(a) and
Fig. 1(c) ($1\gamma\times2\gamma$), and Fig. 1(b) and Fig. 1(c)
($Z\times2\gamma$), respectively, while their sum ($2\gamma$), is
given by the dot-dashed curves. The solid lines represent the full
results of our calculation which include both TPE and $\gamma
Z$-exchange corrections.} \label{Fig_delta}
\end{figure}

In Fig. 2, we show the TPE and $\gamma Z$-exchange corrections to
$A_{PV}$ by plotting $\delta$, defined by
\begin{equation}A_{PV}(1\gamma+Z+2\gamma+\gamma
Z)=A_{PV}(1\gamma+Z)(1+\delta),\end{equation} {\it vs.} $\epsilon
\equiv [1+2(1+\tau)\tan^2\theta_{Lab}/2]^{-1}$, where $\theta_{Lab}$
is the laboratory scattering angle and $\tau=Q^2/4M^2$, at four
different values of $Q^2= 0.1,\,0.5,\, 1.0,\,  5.0 \,  GeV^2$.
$A_{PV}(1\gamma+Z)$ denotes the parity-violating asymmetry arising
from the interference between OPE and $Z$-boson-exchange, i.e.,
Figs. 1(a) and 1(b) while $A_{PV}(1\gamma+Z+2\gamma+\gamma Z)$
includes the effects of TPE and $\gamma Z$-exchange. The full
results are represented by solid curves. To compare our results with
those obtained in the partonic calculation of Ref.
\cite{Afanasev05}, we also show in Fig. 2 the interferences between
OPE and TPE ($1\gamma\times2\gamma$), by dotted lines, as well as
that between $Z$-exchange and TPE ($Z\times2\gamma$) in dashed
lines, with their sum ($2\gamma$) denoted by dot-dashed lines. One
sees that our results at $Q^2 = 5\, GeV^2$, are qualitatively
similar to those obtained in \cite{Afanasev05}, namely,  $1\gamma
\times2\gamma$ contribution cancels that of $Z\times2\gamma$ and
hence their sums are always small (less than 1\%). The $\gamma
Z$-exchange contributions are seen to be much greater than the TPE
effect. In the kinematical regions of HAPPEX and A4 experiments,
i.e., $Q^2\le 0.5 \, GeV^2$  and $\epsilon \ge 0.83$, $\gamma
Z$-exchange effects completely dominate the TPE effects.

To be more quantitative, at $Q^2$=$0.1 \,GeV^2$ the full
correction reaches about 1.9\% at backward angle $135^o$ (SAMPLE),
about 1.4\% at forward angle $35^o$ (A4) and about 0.36\% at
forward angle $6^o$ (HAPPEX). On the other hand, as $Q^2$ grows
larger than $1.0 \,GeV^2$, the full correction decreases and
becomes less than 1\% at forward angles and around 1.5\% at
backward angles.

We now turn to examine the effects of the TPE and $\gamma
Z$-exchange on the values of strange form factors extracted from
HAPPEX \cite{HAPPEX} and A4 \cite{A4} experiments. The parity
asymmetry is conventionally expressed in the following form
\cite{Musolf92},
\begin{eqnarray}
&&A_{PV}(\rho,\kappa)=A_1+A_2+A_3, \nonumber\\
&& A_{1}= -a\rho\left[(1-4\kappa\sin^{2}\theta_{W})-\frac{\epsilon
G^{\gamma,p}_{E}G^{\gamma,n}_{E}
+\tau G^{\gamma,p}_{M}G^{\gamma,n}_{M}}{\epsilon(G^{\gamma,p}_ {E})^2+\tau(G^{\gamma,p}_{M})^2}\right],\nonumber \\
&& A_{2}= a\rho\frac{\epsilon G^{\gamma,p}_{E}G^{s}_{E}
+\tau G^{\gamma,p}_{M}G^{s}_{M}}{\epsilon(G^{\gamma,p}_{E})^2+\tau(G^{\gamma,p}_{M})^2},\nonumber \\
&&A_{3}=a(1-4\sin^{2}\theta_{W})\frac{\epsilon'
G^{\gamma,p}_{M}G_{A}^{Z}}
{\epsilon(G^{\gamma,p}_{E})^2+\tau(G^{\gamma,p}_{M})^2},
\label{A123}
\end{eqnarray}
where $a= G_{F}Q^2/4\pi\alpha\sqrt{2}$, $\epsilon'=\sqrt{\tau
(1+\tau) (1-\epsilon^2)}$, and $\alpha$ the fine structure
constant. When the parameters $\rho$ and $\kappa$ are set to equal
one, Eq. (\ref{A123}) reduces to the expression obtained in tree
approximation. The linear combination of the strange form factors
$G^{s}_{E}+\beta G^{s}_{M}$, with $\beta=\tau G^{\gamma,
p}_{M}/\epsilon G^{\gamma,p}_{E}$ has been extracted from $A_2$ in
Eq. (\ref{A123}).

The latest PDG values \cite{PDG2006} for $\rho$ and  $\kappa$ are
$\rho=0.9876, \kappa=1.0026$. They deviate from one because
higher-order contributions like vertex corrections, corrections to
the propagators and $\gamma Z$-exchange are taken into account.
The effect of the $\gamma Z$ box diagram was estimated in
\cite{Marciano83} for the case of zero momentum transfer $Q\equiv
0$ and gives a contribution of $\Delta \rho$=$-3.7\times 10^{-3}$
and $\Delta\kappa$=$-5.3\times 10^{-3}$ if the onset scale is set
to be $1 GeV$. To avoid double counting one should then subtract
$\Delta \rho$ and $\Delta \kappa$ from $\rho$ and $\kappa$ and use
$\rho'=\rho-\Delta \rho$ and $\kappa'=\kappa-\Delta \kappa$ in Eq.
(\ref{A123}) instead. Consequently, we will set the experimental
parity asymmetry $A_{PV}^{(Exp)}$
\begin{eqnarray}
A_{PV}^{(Exp)}&\equiv &A_{PV}(1\gamma+Z+2\gamma+\gamma Z),
\nonumber\\&=&A_{PV}(\rho',\kappa')(1+\delta). \label{A-corrected}
\end{eqnarray}
With the value we obtain for $\delta$,   we  can then determine
$A_{PV}(\rho',\kappa')$ and extract strange form factors from the
resultant $A_2$. We introduce
\begin{equation}\overline{G}_E^s+\beta\overline{G}_M^s=(G_E^s+\beta
G_M^s)(1+\delta_G), \end{equation}
 to quantify the effect of the
TPE and $\gamma Z$-exchange effect on the extracted values of
strange form factors, where $G_E^s+\beta G_M^s$ and
$\overline{G}_E^s+\beta\overline{G}_M^s$ denote those extracted
from  $A_{PV}(\rho,\kappa)$ and $A_{PV}(\rho',\kappa')$,
respectively. From Eq. (\ref{A123}) we then obtain
\begin{equation}
\delta_G=\frac{A^{Exp}_{PV}(\frac{\Delta\rho}{\rho}-\delta)+4a\rho\sin^{2}\theta_{W}
\Delta\kappa-A_{3}\frac{\Delta\rho}{\rho} }{A^{Exp}_{PV}-A_{0}},
\label{deltaG}
\end{equation}
where $A_{0}=A_{1}+A_{3}$. Note that in HAPPEX and A4 experiments,
the values of $A_{PV}$ are negative while the values of $A_{1}$
and $A_{3}$ are both negative and $A_2$ are positive.

We present our results for $\delta_G$ in Table 1 for HAPPEX and A4
experiments. We see that at forward angles, even though $\delta$,
the corrections to $A_{PV}$, are at most around $1\%$, the
corrections to $G_E^s+\beta G_M^s$, $\delta_G$, are large and
negative and can reach as negative as $-40\%$. We find it is
dominated by the second term of Eq. (\ref{deltaG}). This is because
that though $\Delta \kappa$ is small, its coefficient  is very
large.

\begin{figure}[t]
\centerline{\epsfxsize 3.0 truein\epsfbox{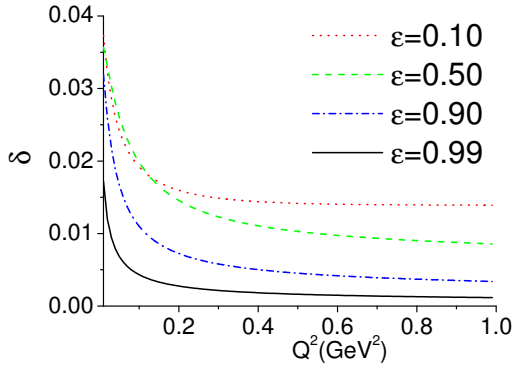}} \caption{
Full TPE and $\gamma Z$-exchange corrections to parity-violating
asymmetry as a function of $Q^2$ from 0.01 to 1 at $\epsilon=0.1,
\,0.5, \,0.9, \,0.99$.}\label{delta-Q2}
\end{figure}

Our results of large $\delta_G$ can be understood by looking at
the $Q^2$ evolution of $\delta$,  depicted in Fig. \ref{delta-Q2}.
Previous estimate of the $\gamma Z$ box diagrams of Fig 1. (d)
\cite{Marciano83} considered only the case of vanishing momentum
transfer between initial and final electrons, corresponding to
$Q^2\equiv 0$ and $\epsilon=1$. It is clear from Fig.
\ref{delta-Q2}
 that the combined effects of the TPE and $\gamma Z$-exchange effect
drops rapidly in the region of $0<Q^2<0.1 GeV^{2}$. In addition,
it drops faster at larger $\epsilon$ than at small $\epsilon$.
Hence the use of the results of \cite{Marciano83} for any finite
$Q^2$ value would grossly overestimate the $\gamma Z$-exchange
effect. However, $\delta(G^{s}_{E}+\beta G^{s}_{M})$ is still
smaller than the current experimental errors even $\delta_G$ is
large. It is because the extracted values of $ G^{s}_{E}+\beta
G^{s}_{M}$ are very small.

\begin{table}[htbp]
\begin{tabular}
{|c|c|c|c|c|c|}
\hline  & I & II & III & IV & V \\
\hline $Q^2(GeV^2)$& 0.477 & 0.1   & 0.109   & 0.23     & 0.108\\
\hline $\epsilon$  & 0.974 & 0.994 & 0.994   & 0.83      & 0.83 \\
\hline $\delta(\%)$  & 0.25 & 0.36  & 0.34 & 0.86  & 1.3 \\
\hline $\delta_G(\%)$  & -22.00 & -12.30  & -39.75 & -3.95 & -3.5 \\
\hline
\end{tabular}
\caption{The corrections $\delta_G$ to $G_E^s+\beta G_M^s$ for
HAPPEX and A4 experiments. I,II and III refer to HAPPEX data in
2004, 2006, and 2007 \cite{HAPPEX}, and IV and V correspond to A4
data in 2004 and 2005, respectively \cite{A4}.} \label{tab1}
\end{table}

In summary, we estimate both the TPE and $\gamma Z$-exchange
corrections to the parity-violating asymmetry of the elastic
polarized electron-proton scattering in a hadronic model. We find
both can reach a few percent in a wide range of momentum transfer,
and are comparable in size to the current experimental measurements
of strange-quark contributions in the proton neutral weak current.
The effect of  $\gamma Z$-exchange is seen in general to be larger
than that of TPE, especially in low $Q^2$ region. Their combined
effects on the extracted values of $G^s_E+\beta G^s_M$ is
surprisingly large, up to $-40\%$ in recent HAPPEX and A4
experiments. The reason is because previous estimate of $\gamma
Z$-exchange effects, as used in current experimental analyses, was
made at $Q^2=0$ and greatly overestimates them at nonvanishing $Q^2$
region.

This work is partially supported by the National Science Council
of Taiwan under grants nos. NSC095-2112-M022-025 (H.Q.Z. and
S.N.Y.) and NSC095-2112-M033-014 (C.W.K.).


\begin{thebibliography}{99}
\bibitem[*]{}Address after August, 2007: Department of Physics,
Southeast University, Nanjing, China

\bibitem{Beck01}D.H. Beck and R.D. McKeown, Annu. Rev. Nucl. Part.
Sci. {\bf 51}, 189 (2001).

\bibitem{Amsler98}C. Amsler, Rev. Mod. Phys. {\bf 70}, 1293
(1998); J. Ellis, Nucl. Phys. A {\bf 684}, 53c (2001).

\bibitem{Titov97}A.I. Titov, Y. Oh, and S.N. Yang, \ Phys. Rev.\ Lett. \
{\bf 79},\ 1634\ (1997), Nucl. Phys. A {\bf 618}, 259 (1997).

\bibitem{Kaplan}
D.\ Kaplan, A.\ Manohar,\ Nucl.\ Phys. \ {\bf B 310},\ 527\
(1988).



\bibitem{SAMPLE}
B. Mueller {\it et al.}, Phys. Rev. Lett. {\bf 78}, 3824 (1997);
D.T. Spayde {\it et al.}, Phys. Lett. {\bf B 583}, 79 (2004).


\bibitem{HAPPEX}
K.A. Aniol {\it et al.} (HAPPEX), Phys. Rev. {\bf C 69}, 065501
(2004), Phys. Lett. {\bf B 635}, 275 (2006); A. Acha {\it et al.}
(HAPPEX), Phys. Rev. Lett. {\bf 98}, 032301 (2007).

\bibitem{A4}
F.E. Maas {\it et al.} (A4), Phys. Rev. Lett. {\bf 93}, 022002
(2004); Phys. Rev. Lett. {\bf 94}, 152001 (2005); B. Glaser (for
the A4 collaboration) Eur. Phys. J. {\bf A 24, S2},141(2005).

\bibitem{G0}
D.S. Armstrong {\it et al.} (G0), Phys. Rev. Lett. {\bf 95},
092001 (2005); C. Furget (for the G0 collaboration), Nucl. Phys.
Proc. Suppl. {\bf 159}, 121 (2006).



\bibitem{Marciano83}
W.J. Marciano and A. Sirlin, Phys.\ Rev. {\bf D 27}, 27 (1983);
Phys.\ Rev. {\bf D 29}, 75 (1984).

\bibitem{Musolf92}
M.J. Musolf and T.W. Donnelly, Nucl. Phys. {\bf A 546}, 509 (1992) ;
M.J. Musolf, {\it et al.}, Phys. Rep. {\bf 239} 1 (1994), Phys. Rev.
{\bf C 60}, 015501 (1999).

\bibitem{Afanasev05}
A.V.\ Afanasev, C.E.\ Carlson,\ Phys.\ Rev. \ Lett.\ {\bf 94},\
212301\ (2005).

\bibitem{Chen04}
Y.C. Chen, A.V. Afanasev, S.J. Brodsky, C.E. Carlson, M.
Vanderhaeghen, Phys. Rev. Lett {\bf 93}, 122301 (2004).

\bibitem{Jone00}
M.K. Jones {\it et al.}, Phys. Rev. Lett. {\bf 84}, 1398 (2000);
O. Gayou {\it et al.}, Phys. Rev. Lett. {\bf 88}, 092301 (2002).

\bibitem{Blunden03} P.G. Blunden, W. Melnitchouk, and J.A. Tjon, Phys. Rev. Lett. {\bf
91}, 142304 (2003).

\bibitem{MoTsai}
L.W. Mo and Y.S. Tsai,
Rev. Mod. Phys. {\bf 41}, 205 (1969).


\bibitem{Beise}
E.J. Beise, M.L. Pitt , D.T. Spayde, Prog. Part. Nucl. Phys. {\bf
54}, 289 (2005).

\bibitem{Feyncalc}
R. Mertig, M. Bohm, and A. Denner, Comput. Phys. Commun. {\bf 64},
345 (1991).

\bibitem{Looptools} T.\ Hahn, M.\ Perez-Victoria, Comput. Phys. Commun. {\bf 118}, 153 (1999).

\bibitem{PDG2006}W.-M. Yao, J. Phys. G {\bf 69}, 1 (2006).

\end{thebibliography}
\end{document}